\documentstyle[12pt]{article} 

\author{H.Falomir$^1$ 
\and 
R.E.Gamboa Sarav\'\i $^1$ 
\and 
M.A.Muschietti$^2$ 
\and
E.M.Santangelo$^1$ 
\and 
J.E.Solomin$^2$ \\ \hfill
\and
{\Large Facultad de Ciencias Exactas, UNLP}\\
$^1$Departamento de F\'{\i}sica\\ 
$^2$Departamento de Matem\'atica} 
\title{Determinants of  Dirac operators with
local boundary conditions. 
\thanks{Partially supported by CONICET, Argentina. } }    
\date{ }

\def\dfrac#1#2{{\displaystyle {#1 \over #2}}}

\def\QATOP#1#2{{#1 \atop #2}}
\def\QDATOP#1#2{{\displaystyle {#1 \atop #2}}}
  
\catcode`\@=11
\def\qed{\ifhmode\unskip\nobreak\fi\ifmmode\ifinner\else\hskip5\p@\fi\fi
 \hbox{\hskip5\p@\vrule width4\p@ height6\p@ depth1.5\p@\hskip\p@}}

\begin{document}

\maketitle 
\begin{abstract} 

We study functional determinants for Dirac operators on manifolds with
boundary. We give, for local boundary conditions, an explicit formula
relating these determinants to the corresponding Green functions. We
finally apply this result to the case of a bidimensional disk under
bag-like conditions. 

\end{abstract} 
                   
\bigskip
\bigskip
% e-mail address: quique@dartagnan.fisica.unlp.edu.ar\newline
PACS number: 03.65.Db \\ 
% Running title: {\bf Determinants of Dirac Operators ...}
                                   
\eject

\section*{I Introduction} 
 
It is well known that functional determinants have wide application in
Quantum and Statistical Physics. Typically, one faces the necessity of
defining a regularized determinant for elliptic differential operators. In
this context, the Dirac first order differential operator plays a central
role. 
 
Seeley's construction of complex powers of elliptic differential operators
provides a powerful tool to regularize such determinants: the so called
$\zeta$-function method \cite{Haw}. 
 
In the case of boundaryless manifolds, this construction has been largely
studied and applied (see, for instance, \cite{Annals} and references
therein). 
 
For manifolds with boundary, the study of complex powers was performed in
\cite{Seeley1,Seeley2} for the case of local boundary conditions, while
for the case of nonlocal conditions, this task is still in progress (see,
for example, \cite{SG}.)
 
In general, the regularized determinant turns out to be nonlocal and so,
it cannot be expressed in terms of just a finite number of Seeley's
coefficients. However, such determinant can always be obtained from the
Green function in a finite number of steps involving these coefficients. 
For boundaryless manifolds this was proved in \cite{Jour.1}, while for a
particular type of local boundary conditions the procedure was introduced
in \cite{Jour.2}. 
 
The aim of this paper is to give the explicit relationship between
determinants and the corresponding Green functions of Dirac operators
under general local elliptic boundary conditions. 
 
Dirac operators defined on manifolds with boundaries have been the subject
of a vast literature (see, for instance, \cite{Moreno, NT} and references
therein), mainly concerning anomalies and index theorems. But, in these
papers, the emphasis was put on nonlocal boundary conditions of the type
introduced in \cite{APS}. We leave for a forthcoming publication the
treatment of such conditions.

The outline of this paper is as follows: 
 
In Section II we introduce some general definitions and conventions,
concerning elliptic boundary problems for Dirac operators. 
 
In Section III we present a formula relating the determinant of the Dirac
operator with its Green function, for the case of local boundary
conditions. 

In Section IV, an explicit computation of the determinant of a Dirac
operator in a bidimensional disk with bag-like boundary conditions is
performed, making use of the results in section III. 
 
\section*{II Elliptic boundary problems, complex powers and regularized
determinants}
 
Throughout this paper we will be concerned with boundary value problems
associated to first order elliptic operators \begin{equation}
\label{OP}D:C^\infty (M,E)\rightarrow C^\infty (M,F), \end{equation} where
$M$ is a bounded closed domain in ${\bf R}^\nu $
 with smooth boundary $\partial M$, and E and $F$ are $k$  
-dimensional complex vector bundles over $M.$ 
 
In a collar neighborhood of $\partial M$ in $M,$ we will take coordinates
$ \bar x=(x,t)$, with $t$ the inward normal coordinate and $x$ local
coordinates for $\partial M$ (that is, $t>0$ for points in $M$ $\setminus
$$\partial M$ and $t=0$ on $\partial M$ ), and conjugated variables
$\bar \xi =(\xi ,\tau )$. 
 
As stated in the Introduction, we will mainly consider the Euclidean Dirac
operator. Let us recall that the free Euclidean Dirac operator $i\not \!
\partial $ is defined as \begin{equation} i\not \! \partial =\sum_{\mu
=0}^{\nu -1}i\gamma _\mu \frac \partial {\partial x_\mu }, \end{equation}
where the matrices $\gamma _\mu $ satisfy \begin{equation} \gamma _\mu
\gamma _\alpha +\gamma _\alpha \gamma _\mu =2\delta _{\mu \alpha },
\end{equation} and that, given a gauge potential $A=\{A_\mu ,\ \mu
=0,...,\nu -1\}$ on $M$, the coupled Dirac operator is defined as
\begin{equation} \label{culo}D(A)=i\not \! \partial +\not \! \! A
\end{equation} with $\not \!\! A=\sum\limits_{\mu =0}^{\nu -1}\gamma _\mu
A_\mu .$

One of the most suitable tools for studying boundary problems is the
Calder\'on projector $Q$ \cite{LibroAmarillo, pelotas}. For the case we
are interested in, $D$ of order 1 as in (\ref{OP}), $Q$ is a (not
necessarily orthogonal) projection from [$L^2(\partial M,E_{/\partial
M})]$ onto the subspace $\{(T\varphi \ /\varphi \in \ker (D)\},$ being
$T:C^\infty (M,E)\rightarrow C^\infty (\partial M,E_{/\partial M})$ the
trace map. 
 
As shown in \cite{LibroAmarillo}, $Q$ is a zero-th order
pseudodifferential operator and its principal symbol $q(x;\xi ),$ that
depends only on the principal symbol of $D,\sigma _1(D)=a_1(x,t;\xi ,\tau
),$ turns out to be the $k\times k$ matrix \begin{equation}
\label{q1}q(x;\xi )=\frac 1{2\pi i}\int_\Gamma \left( a_1^{-1}(x,0;0,1)\
a_1(x,0;\xi ,0)-z\right) ^{-1}dz, \end{equation} where $\Gamma $ is any
simple closed contour oriented clockwise and enclosing all poles of the
integrand in $Im(z)<0.$
 
Q is not unique, since it can be constructed from any fundamental solution
of $D$, but its principal symbol $q(x;\xi )$ is uniquely determined \cite
{LibroAmarillo}.

According to Calder\'on \cite{LibroAmarillo} and Seeley \cite{pelotas},
elliptic boundary conditions can be defined in terms of $q(x;\xi )$. 

\bigskip

{\bf Definition} 1: 
 
Let us assume that the $rank$ of $q(x;\xi )$ is a constant $r$ ( as is
always the case for $\nu \geq 3$ \cite{LibroAmarillo}). 
 
A zero order pseudodifferential operator $B:[L^2(\partial M,E_{/\partial
M})]\rightarrow $ $[L^2(\partial M,G)],$ with $G$ an $r$ dimensional
complex vector bundle over $\partial M,$ gives rise to an {\it elliptic
boundary condition }for a first order operator $D$ as in (\ref{OP}) if,
$\forall \xi :\vert \xi \vert \geq 1,$ \begin{equation}
\label{erre}rank(b(x;\xi )\ q(x;\xi ))=rank(q(x;\xi ))=r\ , \end{equation}
where $b(x;\xi )$ coincides with the principal symbol of $B$ for $\vert
\xi \vert \geq 1.$
 
In this case we say that 
 
\begin{equation} \label{BoundaryProblem}\left\{ \begin{array}{c} D\varphi
=\chi\ \ \rm{ in }\ M \\ \\ BT\varphi =f\ \rm{ on }\ \partial M
\end{array} \right.  \end{equation} is an {\it elliptic boundary problem},
and denote by $D_B$ the closure of $D$ acting on the sections $\varphi $
$\in C^\infty (M,E)$ satisfying $ B(T\varphi )=0.$

An elliptic boundary problem as (\ref{BoundaryProblem}) has a solution $
\varphi \in H^1(M,E)$ for any $(\chi ,f)$ in a subspace of $L^2(M,E)\times
H^{1/2}(\partial M,G)$ of finite codimension. Moreover, this solution is
unique up to a finite dimensional kernel \cite{LibroAmarillo}. In other
words, the operator \begin{equation} (D,BT):H^1(M,E)\rightarrow
L^2(M,E)\times H^{1/2}(\partial M,G) \end{equation} is Fredholm.

When $B$ is a local operator, Definition 1 yields the classical local
elliptic boundary conditions, also called Lopatinsky-Shapiro conditions
(see for instance \cite{Hor}) . 
 
For Euclidean Dirac operators on ${\bf R}^\nu ,$ $E_{/\partial M}=\partial
M\times {\bf C}^k$, local boundary conditions arise when the action of $ B
$ is given by the multiplication by a $\frac k2\times k$ matrix of
functions defined on $\partial M.$
 
Owing to {\it topological obstructions, }chiral Dirac operators in even
dimensions, $\cal{D}$, do not admit{\it \ }local elliptic boundary
conditions (see for example \cite{Booss-B}).  Nevertheless, it is easy to
see from Definition 1 that local boundary conditions can be defined for
the full, either free or coupled, Euclidean Dirac operator $$ D(A)=\left(
\begin{array}{cc} 0 & \cal{D}^{\dagger } \\ \cal{D} & 0 \end{array}
\right) $$ on $M.$

\bigskip

We now sketch Seeley's construction of the complex powers of the operator
$D$ under local elliptic boundary condition $B$ \cite{ Seeley1,
Seeley2,Gil}
 
\bigskip
{\bf Definition} 2: 
 
The elliptic boundary problem (\ref{BoundaryProblem}) admits a cone of
Agmon's directions if there is a cone $\Lambda $ in the $\lambda $-complex
plane such that
 
1) $\forall \bar x\in M$, $\forall \bar \xi \neq 0,$ $\Lambda $ contains
no eigenvalues of the matrix $\sigma _1(D)(\bar x,\bar \xi )$ . 
 
2) $\forall \xi :\vert \xi \vert \geq 1,$ $rank(b(x;\xi )\ q(\lambda
)(x;\xi ))=rank(\ q(\lambda )(x;\xi ))$, $\forall \lambda \in \Lambda $,
 
\noindent where $q(\lambda )$ denotes the principal symbol of the
Calder\'on projector $Q(\lambda )$ associated to $D-\lambda I$, with
$\lambda $ included in $\sigma _1(D-\lambda I)$ (i.e. considering $\lambda
$ of degree one in the expansion of $\sigma ($$D-\lambda I)$ in
homogeneous functions ) \cite{ Seeley1,Gil}. 
 
\bigskip\  
 
An expression for $q(\lambda )(x;\xi )$ is obtained from (\ref{q1}): 
 
\begin{equation} \label{qlam}q(\lambda )(x;\xi )=\frac 1{2\pi
i}\int_\Gamma \bigl( a_1^{-1}(x,0;0,1;0)\ a_1(x,0;\xi ,0;\lambda )-z\bigr)
^{-1}dz, \end{equation} where $\ a_1(x,t;\xi ,\tau ;\lambda )=\sigma
_1(D-\lambda I)$ , with $ \lambda $ considered of degree one as stated
above.

Henceforth, we assume the existence of an Agmon's cone $\Lambda $. 
Moreover, we will consider only boundary conditions $B$ giving rise to a
discrete spectrum $sp(D_B).$ Note that this is always the case for
elliptic boundary problems unless $sp(D_B)$ is the whole complex plane
(see, for instance,\cite {Hor})$.$ Now, for $\vert \lambda \vert $ large
enough, $sp(D_B)\cap \Lambda $ is empty, since there is no $\lambda $ in
$sp(\sigma _1(D_B))\cap \Lambda $.  Then, $sp(D_B)\cap \Lambda $ is a
finite set. 

The usual definition of elliptic boundary conditions through ordinary
differential equations in the normal variable can be recovered from Def. 1
by introducing the ``partial symbol" at the boundary \cite{Seeley1}: Let
us write

\begin{equation} \sigma (D- \lambda I ) = a_{0}(x,t;\xi,\tau;\lambda) +
a_{1}(x,t;\xi,\tau;\lambda), \label{a} \end{equation}
 with $a_l$ homogeneous of degree $l$ in $(\bar{\xi} , \lambda)$. We
replace the coefficients of $D$ by their Taylor expansions in powers of
$t$, and group the resulting terms according to their degree of
homogeneity in $(1/t,\xi, -i\partial_{t}, \lambda)$. More precisely, we
set

\begin{equation} a^{(j)} = a^{(j)}(x,t,\xi,-i\partial_{t},\lambda)
=\sum_{l-k=j}^{}{{t^k\over{k !}}
a_{l}^{k}(x,0,\xi,-i\partial_{t},\lambda)}, \end{equation} with
$a_{l}^{k}= \partial_{t}^{k} a_l $. 

Let us denote $\sigma ^{\prime }(D-\lambda I)=
\sum\limits_{j}^{}{a^{(j)}}$ the partial symbol of $D-\lambda I $ at the
boundary. 

Now, condition 2 is equivalent to the following:
 
2') $\forall \lambda \in \Lambda ,$ $\forall x\in \partial M,$ $\forall
g\in {\bf C}^{r}$, the initial value problem  $$ \begin{array}{c} \sigma
_1^{\prime }(D)(x;\xi )\ u(t)=\lambda \ u(t) \\ \\ b(x;\xi )\ u(t)\vert
_{t=0}=g \end{array} $$ has, for each $\xi \neq 0,$ a unique solution
satisfying $\lim \limits_{t\rightarrow \infty }\ u(t)=0$. This is the form
under which this condition is stated in \cite{Seeley1}. 
 
For $\lambda \in \Lambda $ not in $sp(D_B)$, an asymptotic expansion of
the symbol of $R(\lambda )=(D_B-\lambda I)^{-1}$ can be explicitly given
\cite {Seeley1}: 
 
\begin{equation} \label{AE}\sigma (R(\lambda ))\sim \sum_{j=o}^\infty
c_{-1-j}-\sum_{j=o}^\infty d_{-1-j} \end{equation} where the {\it Seeley
coefficients } $c_{-1-j}$ and $d_{-1-j}$ satisfy
 
\begin{equation} \label{7}\sum_{j=o}^1 a_{1-j}\ \ \circ \
\sum_{j=0}^\infty c_{-1-j}=I \end{equation} with $a_{1-j}$ as in
(\ref{a}), $\circ $ denoting the usual composition of homogeneous symbols,
and \begin{equation} \label{9}\left\{ \begin{array}{c} \sigma ^{\prime
}(D-\lambda )\ \circ \ \sum\limits_{j=o}^\infty d_{-1-j}=0 \\ \\ \sigma
^{\prime }(B)\ \circ \ \sum\limits_{j=o}^\infty d_{-1-j}=\sigma (B)\ \circ
\ \ \sum\limits_{j=0}^\infty c_{-1-j}\quad \rm{at}\ \ t=0 \\ \\ \lim
\limits_{t\rightarrow \infty }\ d_{-1-j}=0 .  \end{array} \right.
\end{equation}

Note that condition 2') implies the existence and unicity of the solution
of (\ref{9}).

The coefficients $c_{-1-j}$ $(x,t;\xi ,\tau ;\lambda )$ and $ 
d_{-1-j}(x,t;\xi ,\tau ;\lambda )$ are meromorphic functions of $\lambda $
with poles at those points where $\det [\sigma _1(D-\lambda )(x,t;\xi
,\tau )]$ vanishes. The $c_{-1-j}$'s are homogeneous of degree $-1-j$ in
$(\xi ,\tau ,\lambda )$ ; the \linebreak $d_{-1-j}$ 's are also
homogeneous of degree $-1-j,$ but in ( $\frac 1t,\xi ,\tau ,\lambda )$
\cite{Seeley1}. 

This gives an approximation to $(D_B-\lambda )^{-1}$, a parametrix
constructed as \cite{Seeley1}
 
\begin{equation} \label{100}P_K(\lambda )=\sum_\varphi \psi \left[
\sum_{j=0}^KOp(\theta _2\ c_{-1-j})-\sum_{j=0}^KOp^{\prime }(\theta _{1\
}d_{-1-j})\right] \ \varphi , \end{equation} where $\varphi $ is a
partition of the unity, $\psi \equiv 1$ in $  Supp(\varphi )$, \\
\begin{equation} \begin{array}{c} \theta _2(\xi ,\tau ,\lambda )=\chi
(\vert \xi \vert ^2+\vert \tau \vert ^2+\vert \lambda \vert ^2) \\ \\ \\
\theta _1(\xi ,\lambda )=\chi (\vert \xi \vert ^2+\vert \lambda \vert ^2),
\end{array} \end{equation} \\ with \\ \begin{equation} \chi (t)=\left\{
\begin{array}{cc} 0 & t\leq 1/2 \\ 1 & t\geq 1 \end{array} \right. ,
\end{equation} \\ and \\ \begin{equation} \begin{array}{c} \displaystyle
Op(\sigma )h(x,t)=\int \sigma (x,t;\xi ,\tau )\ \hat h(\xi ,\tau )\
e^{i(x\xi +t\tau )} \dfrac{d\xi }{(2\pi )^{\nu -1}}\ \dfrac{d\tau }{2\pi
}, \\ \\ \\ \displaystyle Op^{\prime }(\sigma )h(x,t)=\int \int \tilde
\sigma (x,t;\xi ,s)\ \tilde h(\xi ,s)\ e^{ix\xi }\dfrac{d\xi }{(2\pi
)^{\nu -1}}\ \dfrac{ds}{2\pi }, \end{array} \end{equation} \\ where $\
\hat h(\xi ,\tau )$ is defined in (\ref{Fourier}) and \begin{equation}
\tilde h(\xi ,s)=\int \ h(x,s)\ e^{-ix\xi }\ dx.\ \end{equation} \\

Moreover, it can be proved from (\ref{AE}) that, for $\lambda \in \Lambda
,$  \begin{equation} \label{SG}\parallel R(\lambda )\parallel _{L^2}\leq
C\vert \lambda \vert ^{-1} \end{equation} with $C$ a constant
\cite{Seeley1,Gil}. 
 
The estimate (\ref{SG}) allows for expressing the complex powers of $D_B$
as \begin{equation} \label{CP}D_B^z=\frac i{2\pi }\int_\Gamma \lambda ^z\
R(\lambda )\ d\lambda \end{equation} for $Re\ z<0$ , where $\Gamma $ is a
closed path lying in $\Lambda $, enclosing the spectrum \linebreak of
$D_B$ \cite{Seeley2}$.$ Note that such a curve $  \Gamma $ always exists
for $sp(D_B)\cap \Lambda $ finite. 
 
For $Re\ z\geq 0$ , one defines  \begin{equation} D_B^z=D^l\circ
D_B^{z-l}\ , \end{equation} for $l$ a positive integer such that $Re\
(z-l)<0$ . 
 
If $Re(z)<-\nu $, the power $D_B^z$ is an integral operator with
continuous kernel $  J_z(x,t;y,s)$ and, consequently, it is trace class. 
As a function of $z$, $Tr(D_B^z)$ can be extended to a meromorphic
function in the whole complex plane {\bf C}, with only simple poles at
$z=j-\nu ,\ j=0,1,2,...$ and vanishing residues when $z=0,1,2,...$
\cite{Seeley2}. Throughout this paper, analytic functions and their
meromorphic extensions will be given the same name. 

 The function $Tr(D_B^z)$ is usually called $\zeta _{(D_B)}(-z)$ because
of its similarity with the classical Riemann $\zeta $-function: if
$\{\lambda _j\}$ are the eigenvalues of $D_B$, $\{\lambda _j^z\}$ are the
eigenvalues of $D_B^z$; so $  Tr(D_B^z)=\sum \lambda _j^z$ when $D_B^z$ is
a trace class operator. 
 
A regularized determinant of $D_B$ can then be defined as \begin{equation}
\label{DR}Det\ (D_B)=\exp [-\frac d{dz}\ Tr\ (D_B^z)]\vert _{z=0} . 
\end{equation}
 
Now, let $D(\alpha )$ be a family of elliptic differential operators on
$M$ sharing their principal symbol and analytically depending on $\alpha
$. Let $B$ give rise to an elliptic boundary condition for all of them, in
such a way that $D(\alpha )_B$ is invertible and the boundary problems
they define have a common Agmon's cone. Then, the variation of
$Det~D(\alpha )_B$ with respect to $\alpha $ is given by (see, for
example, \cite{APS,Forman}) \begin{equation} \label{DD}\frac d{d\alpha
}\ln \ Det~D(\alpha )_B=\frac d{dz}\left[ \ z\ Tr\{\frac d{d\alpha }\left(
D(\alpha )_B\right) \ D(\alpha )_B^{z-1}\}\right] _{z=0} . \end{equation}
Note that, under the assumptions made, $\frac d{d\alpha }\left( D(\alpha
)_B\right) $ is a multiplication operator.

Given $\alpha_0 $ and $\alpha_1 $ , the quotient ${
De{t(D({\alpha}_{1}))}_{B} \over De{t(D({\alpha}_{0}))}_{B}} $ can be
obtained by integrating the variation in (\ref{DD}) along a path from
$\alpha_0 $ to $\alpha_1 $. 

Although $J_z(x,t;x,t;\alpha),$ the kernel of $D(\alpha)_B^z$ evaluated at
the diagonal, can be extended to the whole $z$-complex plane as a
meromorphic function, the r.h.s. in (\ref{DD}) cannot be simply written as
the integral over $M$ of the finite part of \begin{equation} tr\{\frac
d{d\alpha }\left( D(\alpha )_B\right) \ J_{z-1}(x,t;x,t;\alpha)\}
\end{equation} at $z=0$ (where $tr$ means matrix trace). In fact,
$J_{z-1}(x,t;x,t;\alpha)$ is in general non integrable in the variable $t$
near $  \partial M$ for $z\approx 0$. 
 
Nevertheless, an integral expression for the r.h.s. in (\ref{DD}) will be
constructed in Section III, from the integral expression for
$Tr(D(\alpha)_B^{z-1})$ holding in a neighborhood of $z=0$ and obtained in
the following way \cite{Seeley2}: 
 
if $T>0$ is small enough, the function $j_z(x;\alpha)$ defined as
 \begin{equation} \label{T}j_z(x;\alpha)\ =\int_0^TJ_z(x,t;x,t;\alpha)\ dt
\end{equation} for $Re\ z<1-\nu $, admits a meromorphic extension to {\bf
C} as a function of $z$. So, if $V$ is a neighborhood of $ \partial M$
defined by $t<\epsilon $, with $\epsilon $ small enough,
$Tr(D(\alpha)_B^{z-1})$ can be written as the finite part of
\begin{equation} \label{EI} \int_{M/ V}tr\ J_{z-1}(x,t;x,t;\alpha)\
dxdt+\int_{\partial M}tr\ j_{z-1}(x;\alpha)\ dx\ , \end{equation} where a
suitable partition of the unity is understood.

\section*{III Green functions and determinants} 
 
In this section, we will give an expression for $\frac d{d\alpha }\ln \
Det[D(\alpha )_B]$ in terms of $G_B(x,t;y,s;\alpha )$  , the Green
function of $D(\alpha )_B$ (i.e., the kernel of the operator $D(\alpha
)_B^{-1}).$
 
With the notation of the previous Section, (\ref{DD}) can be rewritten as: 
\begin{equation} \label{DDD}\frac d{d\alpha }\ln \ Det~D(\alpha
)_B=\begin{array}{c} \\ F.P. \\ ^{_{z=0}} \end{array}
 \int_Mtr\left[ \frac d{d\alpha }\left( D(\alpha )_B\right) \
J_{-z-1}(x,t; x,t;\alpha )\right] \ d\bar x\ , \end{equation} where the
r.h.s. must be understood as the finite part of the meromorphic extension
of the integral at $z=0$. 

The finite part of $J_{-z-1}(x,t;x,t;\alpha )$ at $z=0$ does not coincide
with the regular part of $G_B(x,t;y,s;\alpha )$ at the diagonal, since the
former is defined through an analytic extension. 
 
However, it can be shown that there exists a relation between them,
involving a finite number of Seeley's coefficients. In fact, for
boundaryless manifolds this problem has been studied in \cite{Jour.1}, by
comparing the iterated \linebreak limits $F.P.\lim \limits_{z\rightarrow
-1}\{\lim \limits_{\bar y\rightarrow \bar x}J_z(x,t;y,s;\alpha)\}$ and
$R.P.\lim \limits_{\bar y\rightarrow \bar x}\{\lim \limits_{z\rightarrow
-1}J_z(x,t;y,s;\alpha)\}=$\linebreak $R.P.\lim \limits_{\bar y\rightarrow
\bar x}G_{B}(x,t;y,s;\alpha).$
 
In the case of manifolds with boundary, the situation is more involved
owing to the fact that the finite part of the extension of
$J_z(x,t;x,t;\alpha)$ at $z=-1$ is not integrable near $\partial M$ $.$ (A
first approach to this problem appears in \cite{Jour.2}). Nevertheless, as
mentioned in Section 2, a meromorphic extension of $  \int_0^T
J_z(x,t;x,t;\alpha)dt,$ with $T$ small enough can be performed and its
finite part at $z=-1$ turns to be integrable in the tangential variables.
A similar result holds, {\it a fortiori}, for $\int_0^T t^n
J_z(x,t;x,t;\alpha)dt,$ with $n=1,2,3...$
 Then, near the boundary, the Taylor expansion of the function $A_\alpha
=\frac d{d\alpha }D(\alpha )_B$ will naturally appear, and the limits to
be compared are $F.P.\lim \limits_{z\rightarrow -1}\{\lim \limits_{\bar
y\rightarrow \bar x}\int_0^Tt^nJ_z(x,t;y,s;\alpha)dt\} $ and $ R.P.\lim
\limits_{\bar y\rightarrow \bar x}\{\lim \limits_{z\rightarrow
-1}\int_0^Tt^nJ_z(x,t;y,s;\alpha)dt\} $ $=R.P.\lim \limits_{\bar
y\rightarrow \bar x}\int_0^Tt^nG_{B}(x,t;y,s;\alpha)dt.$
 
The starting point for this comparison is to carry out asymptotic
expansions and to analyze the terms for which the iterated limits do not
coincide (or do not even exist). 
 
An expansion of $G_B(x,t,y,s)$ in $M\backslash \partial M$ in homogeneous
and logarithmic functions of $(\bar x-\bar y)$ can be obtained from
(\ref{AE}) for $\lambda =0$:  \begin{equation} \label{?} \begin{array}{c}
G_B(x,t,y,s)= \sum_{j=1-\nu }^0h_j(x,t,x-y,t-s)+M(x,t)\log \vert
(x,t)-(y,s)\vert \\ \\ +R(x,t,y,s), \end{array} \end{equation} with $h_j$
the Fourier transform{\cal \ }${\cal F}^{-1}(c_{-\nu -j})$ of $  c_{-\nu
-j}$ for $j>0$ and \linebreak $h_0={\cal \ }{\cal F}^{-1}(c_{-\nu })-\
M(x,t)\log \vert (x,t)-(y,s)\vert .$ The function $M(x,t)$ will be
explicitly defined below (see (\ref{MM})). Our convention for the Fourier
transform is \begin{equation} \label{Fourier} \begin{array}{c}
\displaystyle {\cal F}(f)(\bar \xi )=\hat f(\bar \xi )=\int f(\bar x)\
e^{-i\bar x.\bar \xi }\ d\bar x, \\ \\ \displaystyle {\cal \ }{\cal
F}^{-1}(\hat f)(\bar x)=f(\bar x)=\dfrac 1{(2\pi )^\nu }\int \hat f(\bar
\xi )\ e^{i\bar x.\bar \xi }\ d\bar \xi .  \end{array} \end{equation}
 
For $t>0$, $R(x,t,y,s)$ is continuous even at the diagonal ($y,s)=(x,t)$. 
Nevertheless, $R(x,t,y,s)\vert _{(y,s)=(x,t)}$is not integrable because of
its singularities at $t=0$. On the other hand, the functions
$t^nR(x,t,y,t)$ are integrable with respect to the variable $t$ for $y\neq
x$ and $  n=0,1,2,....$An expansion of $\int_0^\infty t^nR(x,t,y,t)dt$ in
homogeneous and logarithmic functions of $(x-y)$ can also be obtained from
($\ref{AE})$:  \begin{equation} \label{Asn} \int_0^\infty t^nR(x,t,y,t)
dt=\sum_{j=n+2-\nu }^0g_{j,j+n+2-\nu}(x,x-y)+M_n(x)\log (\vert x-y\vert
)+R_n(x,y) \end{equation} where $R_n(x,y)$ is continuous even at $y=x$,
and $g_{j,j+n+2-\nu}$ is the Fourier transform of the (homogeneous
extension of) $\int_0^\infty t^n\tilde d_{-1-j}(x,t,\xi ,t,0)dt$, with
\begin{equation} \label{`d}\tilde d_{-1-j}(x,t,\xi ,s,\lambda
)=-\int_{\Gamma ^{-}}e^{-is\tau }\ d_{-1-j}(x,t,\xi ,\tau ,\lambda )\
d\tau \end{equation} for $\Gamma ^{-}$ a closed path enclosing the poles
of $d_{-1-j}(x,t,\xi ,\tau ,\lambda )$ lying in \linebreak $\{Im\ \tau
>0\}$. 
 
Since $\tilde d_{-1-j}$ is homogeneous of degree $-j$ in (1/t, $\xi
,1/s,\lambda )$, $g_{j,j+n+2-\nu}$ turns out to be homogeneous of degree
$j+n+2-\nu$ in $x-y$ 
. 
 
\bigskip\  
 
From the forementioned comparison, the following Theorem can be shown to
hold ( the proof will be given in \cite{tutorial}): 

{\bf Theorem 1:} {\it Let $M$ be a bounded closed domain in ${\bf R}^\nu $
with smooth boundary $\partial M$ and $E$ a $k$-dimensional complex vector
bundle over $M$. 
 
Let $(D_\alpha )_{B}$ be a family of elliptic differential operators of
first order, acting on the sections of $E$, with a fixed local boundary
condition $B$ on $ \partial M$, and denote by $J_z(x,t;x,t;\alpha)$ the
meromorphic extension of the evaluation at the diagonal of the kernel of
$(D_\alpha )_B^z$. 
 
Let us assume that, for each $\alpha $, $(D_\alpha )_B$ is invertible, the
family is differentiable with respect to $  \alpha ,$ and $\dfrac \partial
{\partial \alpha }(D_\alpha )_{B}f=A_\alpha f$ , with $A_\alpha $ a
differentiable function. 
 
If $V$ is a neighborhood of $\partial M$ defined by $t < \epsilon$ and
$T>0$ small enough, then: 
 
a) } \begin{equation} \label{Cuculiu} \begin{array}{c} \displaystyle
\dfrac \partial {\partial \alpha }\ln \ Det(D_\alpha )_B
 \\  = 
\begin{array}{c} \\ F.P. \\ ^{_{z=-1}} \end{array} \displaystyle \left[
\int_{ \partial M}\int_0^Ttr\left\{ A_\alpha (x,t) \ J_z(x,t;x,t;\alpha)\
\right\} dtdx\right] \\ +\begin{array}{c} \\ F.P. \\ ^{_{z=-1}}
\end{array} \displaystyle \left[ \int_{ M/V}tr\left\{ A_\alpha (\bar x)\
J_{z}(\bar x;\bar x;\alpha)\ \right\}d\bar x\right], \end{array}
\end{equation} {\it where a suitable partition of the unity is understood.
(This expression must be understood as the finite part at $z=-1$ of the
meromorphic extension). 
 
b) For every $\alpha$, the integral $\int_0^TA_\alpha (x,t)\ J_z
(x,t;x,t;\alpha)dt\ $ is a meromorphic function of $z$, for each $x\in
\partial M$, with a simple pole at $z=-1$. Its finite part (dropping, from
now on, the index $\alpha$ for the sake of simplicity) is given by}
\begin{equation} \label{reputamadre} \begin{array}{c} \begin{array}{c} \\
\displaystyle F.P. \\ ^{_{z=-1}} \end{array} \displaystyle \int_0^TA(x,t)\
J_z(x,t;x,t)dt\\

\displaystyle = -\int_0^TA(x,t)\int_{\vert (\xi ,\tau )\vert =1}\frac
i{2\pi }\int_\Gamma \dfrac{\ln \lambda }\lambda \ c_{-\nu }(x,t;\xi ,\tau
;\lambda )\ d\lambda \ \dfrac{d\sigma _{\xi ,\tau }}{{(2\pi )^\nu }}\ dt\
\\ \\ \displaystyle +\sum\limits_{l=0}^{\nu -2} \dfrac{\partial
_t^lA(x,0)}{l!}\int_{\vert \xi \vert =1}\int_0^\infty t^l\ \frac i{2\pi
}\int_\Gamma \dfrac{\ln \lambda }\lambda \ \tilde d_{-(\nu -1)+l}(x,t;\xi
,t;\lambda)\ d\lambda \ \ dt\ \dfrac{d\sigma _\xi }{{{(2\pi )^{\nu -1}}}}\
\\ \\ \displaystyle +\lim \limits_{y\rightarrow x}\left\{
\int_0^TA(x,t)\left[ G_B(x,t;y,t)-\sum\limits_{l=1-\nu
}^0h_l(x,t;x-y,0)\right. \right.  \\ \\ \displaystyle \left. -M(x,t)\
\dfrac{\Omega _\nu }{{(2\pi )^\nu }}\left( \ln \vert x-y\vert ^{-1}+{\cal
K}  _\nu \right) \right] dt \\ \\ \displaystyle +\sum\limits_{j=0}^{^{\nu
-2}}\sum\limits_{l=0}^{\nu -2-j} \dfrac{\partial _t^lA(x,0)}{l!}\
g_{j,l-(\nu -2-j)}(x,x-y)\ \\ \\ \displaystyle +\left. \sum_{l=0}^{^{\nu
-2}}\dfrac{\partial _t^lA(x,0)}{l!}\ M_{\nu -2-l}(x)\ \dfrac{\Omega _{\nu
-1}}{{{(2\pi )^{\nu -1}}}}\left( \ln \vert x-y\vert ^{-1}+{\cal K}_{\nu
-1}\right) \right\} , \end{array} \end{equation} {\it with }
\begin{equation} \label{MM} \begin{array}{c} \displaystyle M(x,t)=\dfrac
1{\Omega _\nu }\int_{\vert (\xi ,\tau )\vert =1}c_{-\nu }(x,t;\xi ,\tau
;0)\ \ d\sigma _{\xi ,\tau } \\ \\ \displaystyle M_j(x)=\dfrac 1{\Omega
_{\nu -1}}\int_{\vert \xi \vert =1}\int_0^\infty t^{\nu -2-j}\ \ \tilde
d_{-1-j}(x,t;\xi ,t;0)\ dt\ d\sigma _\xi , \end{array} \end{equation} {\it
where $\Omega _n=Area(S^{n-1})$, ${\cal K}_\nu =\ln 2-\frac 12\gamma
+\frac 12\frac{\Gamma ^{\prime }(\nu /2)}{\Gamma (\nu /2)}$ with $\gamma $
the Euler's constant and where $h_{l}$ and $g_{l}$ are related to the
Green function $G_B$ as in (\ref{?}) and (\ref{Asn}) } \begin{equation}
\label{hg} \begin{array}{c} \displaystyle h_{1-\nu +j}(x,t;w,u)
 = {\cal F}_{(\xi ,\tau )}^{-1}\ \left[ c_{-1-j}(x,t;(\xi ,\tau 
)/\vert (\xi ,\tau )\vert ;0)\ \vert (\xi ,\tau )\vert ^{-1-j}\right] 
(w,u), \\ \\   
 
\displaystyle h_0(x,t;w,u) = {\cal F}_{(\xi ,\tau )}^{-1}\left[
P.V.\left\{ \left( c_{-\nu }(x,t;(\xi ,\tau )/\vert (\xi ,\tau )\vert
;0)-M(x,t)\right) \ \vert (\xi ,\tau )\vert ^{-\nu }\right\} \right]
(w,u), \\ \\
 
\displaystyle g_{j,l}(x,w)
 = {\cal F}_\xi ^{-1}\left[ \int_0^\infty t^n\ \tilde 
d_{-1-j}(x,t;\xi /\vert \xi \vert  ,t;0)\ dt\vert \xi \vert  
^{-1-j-n}\right] (w), \\  \\ 

with \hfill \\

\centerline{$l = j + n - \nu + 2$,} \\

and \hfill \\

\displaystyle g_{j,0}(x,w) ={\cal F}_\xi ^{-1}\left[ P.V.\left[
\int_0^\infty t^{\nu -j-2}\ \tilde d_{-1-j}(x,t;\xi /\vert \xi \vert
,t;0)\ dt-M_j(x)\right] \vert \xi \vert ^{-(\nu -1)}\right] (w). 
\end{array} \end{equation} \\ {\it c) The integral $\int_{M\backslash
V}tr\left[ A(\bar x)\ J_{z}(\bar x;\bar x)\right] \ d\bar x $ in the
second term in the r.h.s. of (\ref{Cuculiu}) , is a meromorphic function
of $z$ with a simple pole at $z=-1$. Its finite part is given by }
\begin{equation} \begin{array}{c} \begin{array}{c} \\ F.P. \\ ^{_{z=-1}}
\end{array} \displaystyle \int_{M\backslash V}tr\left[ A(\bar x)\
J_{z}(\bar x;\bar x)\right] \ d\bar x \\ \displaystyle =\int_{M\backslash
V}A(\bar x)\int_{\vert \bar \xi \vert =1}\dfrac i{2\pi }\int \dfrac{\ln
\lambda }\lambda \ c_{-\nu}(\bar x,\bar \xi ;\lambda )\ d\lambda
\dfrac{d\bar \xi }{(2\pi )^\nu } \\ \\ \displaystyle +\int_{M\backslash
V}\lim \limits_{\bar y\rightarrow \bar x}\ A(\bar x)[G_B(\bar x,\bar
y)-\sum\limits_{l=1-\nu }^0h_l(\bar x,\bar x-\bar y) \\ \\ \displaystyle
-M(\bar x)\dfrac{\Omega _\nu }{(2\pi )^\nu }(\ln \vert \bar x-\bar y\vert
^{-1}+{\cal K}_\nu )]\ d\bar x.  \end{array} \end{equation}

\bigskip
 
\bigskip 

This Theorem gives a closed expresion for the evaluation of the
determinant when the associated Green function is known. 

Awful as it looks, (\ref{reputamadre}) is not so complicated: In the first
place, all terms can be systematically evaluated. Moreover, the terms
containing $h_l$ subtract the singular part of the Green function in the
interior of the manifold (see (\ref{?})) and can, thus, be easily
identified from the knowledge of $G_B$. $R(x,t,y,t)$, the regular part so
obtained, is still nonintegrable near the boundary. Those terms containing
$g_{j,l}$ subtract the singular part of the integrals $\int_0^T\ t^n\
R(x,t,y,t)\ dt$ (see (\ref{Asn})). Finally, the terms containing $c_{-\nu
}$ and $d_{-\nu +1} $ arise as a consequence of having replaced an
analytic regularization by a {\it point splitting } one. 
 
Even though Seeley's coefficients $c$ and $\tilde d$ are to be obtained
through an iterative procedure, which can make their evaluation a tedious
task, in the cases of physical interest only the few first of them are
needed.  In fact, for the two dimensional example in the next section we
will only need two such coefficients.   

%\eject

\section*{IV Two dimensional Dirac operator on a disk.}

In this section, we will use the method previously discussed to evaluate
the determinant of the operator $D=\ \not \!\!\!i\partial +\not \!\!A$
acting on functions defined on a two dimensional disk of radius $R.$ A
family of local bag-like \cite{Bag} elliptic boundary conditions will be
assumed. 
 
We take $A_\mu $ to be an Abelian field in the Lorentz gauge; as it is
 well known, it can be written as $A_\mu =\epsilon _{\mu \nu }\ \partial
_\nu \phi \ (\epsilon _{01}=-\epsilon _{10}=1)$. For $\phi $ we choose a
smooth bounded function $\phi =\phi (r)$ .  Notice that, with these
assumptions, $A_r=0$ and $A_\theta (r)=-\partial _r\phi (r).$ We call
\begin{equation} \label{3.2}\Phi =\oint_{r=R}\ A_\theta \ R\ d\theta
=-2\pi R\ \partial _r\phi (r) \vert _{r=R}.  \end{equation}

The free Dirac operator in polar coordinates is:  \begin{equation} i\not
\! \partial =i(\gamma _r\ \partial _r+\frac 1r\gamma _\theta \
 \partial _\theta ), \end{equation}
 
with \begin{equation} \label{gampol}\gamma _r=\left( \begin{array}{cc} 0 &
e^{-i\theta } \\ e^{i\theta } & 0 \end{array} \right) ,\qquad \gamma
_\theta =\left( \begin{array}{cc} 0 & -ie^{-i\theta } \\ ie^{i\theta } & 0
\end{array} \right) .  \end{equation}
 
With these conventions, the full Dirac operator can be written as: 
\begin{equation} \label{op}D=e^{-\gamma _5\phi (r)\ }i\not \! \partial \
e^{-\gamma _5 \phi (r)}.  \end{equation}

\bigskip

Now, in order to perform our calculations, we consider the family of
operators:  \begin{equation} \label{opalf}D_\alpha =i\not \! \partial
+\alpha \not \! \! A=e^{-\alpha
 \gamma _5\phi (r)\;\ }i\not \! \partial \ e^{-\alpha \gamma _5\phi (r)},\
\rm{with }\ 0\leq \alpha \leq 1, \end{equation} which will allow us to go
smoothly from the free to the full Dirac operator.  If we call
\begin{equation} W(\alpha )=\ln \ Det(D_\alpha )_B, \end{equation} where
$B$ represents the elliptic boundary condition, we have \begin{equation}
\dfrac \partial {\partial \alpha }W(\alpha )=\begin{array}{c} \\ F.P. \\
^{_{z=0}} \end{array} \left[ Tr\left( \not\!\! A\ (D_\alpha
)_B^{-z-1}\right) \right].  \end{equation} From the Theorem in the
previous section we get:  \begin{equation} \label{putamadre}
\begin{array}{c} \displaystyle \dfrac \partial {\partial \alpha }W(\alpha
)= \frac 1{(2\pi )^2}\ tr\left\{ \int \ \lim \limits_{y\rightarrow
x}\left[ \int \left[ \not \! \! A(t)\left( 4\pi ^2G_{B}(x,t,y,t)\right.
\right. \right.  \right.  \\ \\ \displaystyle -\frac 1{\vert x-y\vert
}\int\ e^{i\xi \frac{(x-y)}{\vert x-y\vert }}\ c_{-1}(x,t;\frac{(\xi ,\tau
)}{\vert (\xi ,\tau )\vert };0) \ d\xi \ d\tau \ \\ \\ \displaystyle
-\int_{\vert (\xi ,\tau )\vert \geq 1}\ e^{i\xi (x-y)}\ c_{-2}(x,t;\xi
,\tau ;0)\ d\xi \ d\tau \\ \\ \displaystyle \biggl. -\int\frac i{2\pi
}\int_\Gamma \ \frac{\ln \lambda}\lambda \ c_{-2}(x,t;\frac{(\xi ,\tau
)}{  \vert (\xi ,\tau )\vert };\lambda )\ d\lambda\
d\sigma_{\xi,\tau}\biggr)\\ \\ \displaystyle + \not \! \! A(0)\ \biggl(
\int_{\vert \xi \vert \geq 1}\ e^{i\xi (x-y)}\ \tilde d_{-1}(x,t;\xi
,t;0)\ d\xi\biggr. \\ \\ \displaystyle \left. \left. \biggl. \biggl. +\
\int \frac i{2\pi }\int_\Gamma \frac{\ln \lambda}\lambda \ \tilde
d_{-1}(x,t;  \frac{(\xi )}{\vert \xi \vert },t;\lambda )\ d\lambda\
d\sigma_\xi \biggr)
 \biggr]dt \right]dx \right\} , \end{array} \end{equation} where the
Fourier transforms of $c_{-2}$ and $\tilde d_{-1}$ have been left
explicitly indicated. 
 
Now, the coefficients $c$ and $\tilde d$ in the previous equation are
those appearing in the asymptotic expansion of the resolvent $(D_\alpha -
\lambda I$ $)^{-1}$. 
 
From (\ref{op}), the symbol of $(D_\alpha -\lambda I)$ is: 
\begin{equation} \label{sim} \begin{array}{c} \sigma (D_\alpha -\lambda
I)=(- \not \xi -\lambda I)+\alpha \not \! \! A \\ \\ \displaystyle
=a_1(\theta ,t,\xi ,\tau ,\lambda )+a_0(\theta ,t,\xi ,\tau
 ,\lambda ), \end{array} \end{equation} where \begin{equation}
\begin{array}{c} a_1=- \not \xi -\lambda I, \\ \\ a_0=\alpha \not \! \! A. 
\end{array} \end{equation} The required Seeley's c-coefficients are given
by \cite{Annals} :  \begin{equation} \label{c} \begin{array}{c}
\displaystyle c_{-1}=\frac 1{(\lambda ^2-\xi ^2-\tau ^2)}( \not \! \xi
-\lambda I), \\ \\ \displaystyle c_{-2}=\frac \alpha {(\lambda ^2-\xi
^2)^2}(2\lambda \xi _\mu
 A_\mu 
I-(\lambda ^2-\xi ^2)\not \! \! A-2\xi _\mu A_\mu \not \! \xi ),  
\end{array} 
\end{equation} 
where $\not \xi =\xi \gamma _\theta +\tau \gamma _t$ . 
 
As regards the boundary contributors to the parametrix, i.e., the
coefficients $d_{-1-j}$ are the solutions of (\ref{9}).  In our case, the
equation to be solved is \begin{equation} \label{recast}(-\lambda I-\xi
\gamma _\theta +i\gamma _t\partial _t)d_{-1}=0, \end{equation} with
boundary conditions \begin{equation} \label{bcd1}b_0\ d_{-1}=b_0\ c_{-1}\
\ \rm{\ at }\ t=0, \end{equation} plus the vanishing of $\ d_{-1}$ as
$t\rightarrow +\infty .\;$(\ref{recast}) can be recast in the form
\begin{equation} \label{eqd1}\partial _td_{-1}=-M\ d_{-1,} \end{equation}
where $M=\xi \gamma _5+i\lambda \gamma _t$ . It can be easily verified
that \begin{equation} \begin{array}{c} tr(M)=0, \\ \\ M^2=(\xi ^2-\lambda
^2)I.  \end{array} \end{equation} So, $M$ has eigenvalues $\pm \sqrt{\xi
^2-\lambda ^2}$, corresponding to the eigenvectors \begin{equation} u_{\pm
}=\left( \begin{array}{c} ie^{-i\theta }(\xi \pm \sqrt{\xi ^2-\lambda ^2})
\\ \lambda \end{array} \right) .  \end{equation} Since $d_{-1}\rightarrow
0$ for $t\rightarrow \infty $ , we get \begin{equation}
\label{d}d_{-1}(x,t;\xi ,\tau ;\lambda )=e^{-t\sqrt{\xi ^2-\lambda ^2} 
}u_{+}\otimes \left( \QATOP{f}{g}\right) ^{\dagger }, \end{equation} where
the vector $\left( \QATOP{f}{g}\right) $ must be determined from
 the 
boundary condition at $t=0~(r=R),$ given by (\ref{bcd1}). 
 
We now consider a parametric family of bag-like local boundary conditions
leading to an elliptic boundary problem, \begin{equation} \label{b0}
b_0=\left( 1,w\ e^{-i\theta }\right) , \end{equation} with $w$ a nonzero
complex constant. (Notice that these boundary conditions reduce to those
of an MIT bag \nobreak \cite{Bag} when $ w=\pm 1$.)
 
We define the operator $(D_\alpha )_B$ as the differential operator in (
\ref{op}) , acting on the dense subspace of functions satisfying
\begin{equation} \label{lbc} B\ \psi \equiv b_0 \psi \vert _{t=0}=0. 
\end{equation} It is easy to verify that this operator has no normalizable
zero modes (Notice that these are not the most general local elliptic
boundary conditions. In fact, zero modes would in general arise if one
allowed $w$ to depend on $\theta $). 

Now, from (\ref{bcd1}) and the expression for $c_{-1}$ given in  (\ref{c}),
 it turns out that:  \begin{equation} \begin{array}{c} \left(
\QDATOP{f}{g}\right) ^{\dagger }=\dfrac{e^{i\theta }}{(\xi ^2+\tau
^2-\lambda ^2) (\lambda w+i\xi +i\sqrt{\xi ^2-\lambda ^2)}}\hfill \\ \\
\qquad \qquad \qquad \qquad \times \left(\ \lambda +w\ (-i\xi +\tau )
\qquad e^{-i\theta }(i\xi +\tau +\lambda \ w)\ \right). \end{array}
\end{equation} Replacing this expression into (\ref{d}), and taking into
account (\ref{`d}), we finally get:  \begin{equation} \label{dm1til}
\begin{array}{c} \tilde d_{-1}=\pi i \dfrac{e^{-(u+t)\sqrt{\xi ^2-\lambda
^2}}}{\sqrt{\xi ^2-\lambda ^2}(iw \lambda -\xi -\sqrt{\xi ^2-\lambda
^2})}\ \ \hfill \\ \\ \times \left( \begin{array}{cc} \scriptstyle (\xi
+\sqrt{\xi ^2-\lambda ^2})\ (i\lambda +w(\xi +\sqrt{ \xi ^2-\lambda ^2} ))
\ \ & \scriptstyle e^{-i\theta }(\xi + \sqrt{\xi ^2-\lambda ^2})\
(iw\lambda -\xi +\sqrt{\xi ^2-\lambda ^2}) \\ \\ \scriptstyle -i\lambda \
e^{i\theta }(i\lambda -w(\xi +\sqrt{\xi ^2- \lambda ^2})) & \scriptstyle
-i\lambda \ (iw\lambda -\xi +\sqrt{\xi ^2-\lambda ^2}) \end{array} \right)
.  \end{array} \end{equation}
 
\bigskip

In order to apply (\ref{putamadre}), we look for the function
 $G_{B}(x,y)$ satisfying:  \begin{equation} \begin{array}{c} D_\alpha \
G_{B}(x,y)=\delta (x,y), \\ \\ B\ G_{B}(x,y)\vert _{x\in \partial \Omega
}=0, \end{array} \end{equation} where $D_\alpha $ and $B,$ are given by
equations (\ref{op}) and
 (\ref{lbc}) respectively. Now, with the notation \begin{equation}
\begin{array}{c} x=(x_0 , x_1)=(r\cos \theta ,r\sin \theta ), \\ X=x_0+i\
x_1=r\ e^{i\theta }, \\ \\ y=(y_0 , y_1)=(\rho \cos \varphi ,\rho \sin
\varphi ),\ \\ Y=y_0+i\ y_1=\rho \ e^{i\varphi }, \end{array}
\end{equation} it is easy to see that $\ G_{B}(x,y)$ is given by
\begin{equation} \label{green1}G_{B}(x,y)=\frac 1{2\pi i}\left(
\begin{array}{cc} \frac{R\ w\ e^{\alpha (\phi (x)+\phi (y)-2\phi
(R))}}{XY^{*}-R^2} & \frac{  e^{\alpha (\phi (x)-\phi (y))}}{X-Y} \\ \\
\frac{e^{-\alpha (\phi (x)-\phi (y))} }{(X-Y)^{*}} & \frac{R\ e^{-\alpha
(\phi (x)+\phi (y)-2\phi (R))}}{w\ (XY^{*}-R^2)^{*}} \end{array} \right) . 
\end{equation}

With these elements at hand, we now perform the calculation of the
determinant . 
 
From (\ref{green1}), one can see that \begin{equation} G_{B}(\theta
,r,\varphi ,r) \begin{array}{c} \\ \sim \\ \scriptstyle \varphi
\rightarrow \theta \end{array} \rm{diagonal\ matrix }+\frac 1{2\pi i\ r\
(\theta -\varphi )}\ \gamma _\theta .  \end{equation}
 
When replaced into (\ref{putamadre}), we get for the first term in the
 r.h.s. 
 
\begin{equation} tr\left\{ A_\theta \ \gamma _\theta \ G_B(\theta
,r,\varphi ,r)\right\} \begin{array}{c} \\ \sim \\ \scriptstyle \varphi
\rightarrow \theta \end{array} \frac{A_\theta }{\pi i\ r\ (\theta -\varphi
)}.  \end{equation}
 
For the second term in (\ref{putamadre}) \begin{equation} -\frac 1{4\pi
^2\vert x-y\vert }\int d\xi \ d\tau \ e^{i\xi \frac{(x-y)} {\vert x-y\vert
}}\ c_{-1}(x,t;\frac{(\xi ,\tau )}{\vert (\xi ,\tau )\vert }; \lambda =0)
\begin{array}{c} \\ \sim \\ \scriptstyle \varphi \rightarrow \theta
\end{array} \frac{-1}{2\pi i\ r\ (\theta -\varphi )}\ \gamma _\theta ,
\end{equation} which exactly cancels the singularity of the Green
function.  Therefore, the contribution of the first two terms in
(\ref{putamadre}) vanishes. 
 
As regards the third term, 
 \begin{equation} \label{w3} \begin{array}{c} \displaystyle
\frac{-1}{(2\pi )^2}\ tr\int \lim \limits_{y\rightarrow x}\ \not\!\! 
A(t)\int_{\vert (\xi ,\tau )\vert \geq 1}e^{i\xi (x-y)}\ c_{-2}(x,t;\xi
,\tau ;0)\ d\xi \ d\tau\ dx\ dt\\ \\ \displaystyle= \frac{-\alpha }{2\pi
^2}\ \lim \limits_{y\rightarrow x}\ \int A_\theta ^2\ d^2x\ \int_{\vert
(\xi ,\tau )\vert \geq 1} e^{i\xi (x-y)\ }\frac{  (\tau ^2-\xi ^2)}{(\xi
^2+\tau ^2)^2}\ d\xi \ d\tau \\ \\ \displaystyle= \frac{-\alpha }\pi \int
A_\theta ^2\ d^2x\ \ \lim \limits_{y\rightarrow x}\int_{\vert x-y\vert
}^\infty J_2(u) \ \ \frac{du}u\ =\ \frac{-\alpha } {2\pi }\int A_\nu \
A_\nu\ d^2x\ .  \end{array} \end{equation} where $J_2(u)$ is the Bessel
function of order two. 
 
Now, the fourth term in ( \ref{putamadre}) is:  \begin{equation}
\label{w4} \begin{array}{c} \displaystyle\frac{-1}{(2\pi )^2}\ tr\int \not
\! \! A(t)\int \frac i{2\pi }\int_\Gamma \ \ln \lambda \ \ c_{-2}(x,t; 
\frac{(\xi ,\tau )}{\vert (\xi ,\tau )\vert};\lambda ) \ \ \frac{d\lambda
} \lambda \ d\sigma_{\xi,\tau}\ dx\ dt\\\ \\ \displaystyle= \frac{-i\alpha
}{4\pi ^3}\int A_\theta ^2\ d^2x\ \int_\Gamma \ \ \frac{\ln \lambda
}{(\lambda ^2-1)^2} \int \ (1-\lambda ^2-2\xi ^2)\ d\sigma_{\xi,\tau}\
\frac{d\lambda } \lambda \ \\ \\ \displaystyle=\frac{i\alpha }{2\pi
^2}\int A_\theta ^2\ d^2x\ \ 2\pi i\ \int_0^\infty \frac{  \mu \ d\mu
}{(\mu ^2+1)^2} \ =\ \frac{-\alpha }{2\pi }\int A_\nu \ A_\nu\ d^2x\ . 
\end{array} \end{equation} This term gives rise to a contribution
identical to that of (\ref{w3}). 
 
The last term in (\ref{putamadre}) is 
 
\begin{equation} \label{last} \begin{array}{c} \displaystyle\frac i{(2\pi
)^3}\ tr\int \not \! \! A(0)\ \sum\limits_{\xi =\pm 1}\int_\Gamma \ \ln
\lambda \ \ \tilde d_{-1}(x,t;\frac \xi {\vert \xi \vert },t;\lambda ) \
\frac{d\lambda }\lambda\ dx\ dt\ \\ \\ \displaystyle=\frac{i\ \Phi }{(2\pi
)^2}\int_\Gamma \ \frac{u\ \ln \lambda }{(1+\ u^2\ \lambda ^2)}[\lambda \
\sqrt{1+u^2}-i \sqrt{1-\lambda ^2}]\ \frac{d\lambda }{\sqrt{1-\lambda ^2}}
, \end{array} \end{equation} where $u=(1-w^2)\ /\ 2w.$ We choose the curve
$\Gamma $ as in Fig. 
 \ref
{figura1}.

\begin{figure} \begin{center} \begin{picture}(150,150)(-75,-50) \put
(0,-50){\line(0,1){150}} \put (-75,0){\line(1,0){150}} \put
(5,100){\vector(0,-1){60}} \put (5,40){\line(0,-1){30}} \put
(5,0){\oval(20,20)[r]} \put (5,-10){\line(-1,0){10}} \put
(-5,0){\oval(20,20)[l]} \put (-5,10){\vector(0,1){60}} \put
(-5,70){\line(0,1){30}} \put (20,20){\makebox(0,0){$\Gamma$}}

\end{picture}
\end{center}
\caption{ The contour $\Gamma$ }
\label{figura1}
\end{figure}

Therefore, (\ref{last}) reads \begin{equation} \label{w6} \begin{array}{c}
\displaystyle -\dfrac \Phi {2\pi }\ u\ \int_0^\infty \dfrac 1{(1-\ u^2\
 \mu ^2)}\left[ \mu \ \ \dfrac{\sqrt{\ 1\ +\ u^2}}{\sqrt{\ 1\ +\ \mu ^2}}\
-\ 1\right] d\mu \ \\
 \\  
=\dfrac{-\Phi }{4\pi }\ \ln \ w^2.  \end{array} \end{equation}
 
Putting all pieces together ((\ref{w3}), (\ref{w4}) and (\ref{w6}), we
finally find:  \begin{equation} \begin{array}{c} \displaystyle\ln Det
(D)_B-\ln Det (\!i\not\!\partial )_B=-\frac 1{2\pi }
 \int_\Omega A_\nu \ A_\nu \ d^2x\ \ 
-\frac \Phi {4\pi }\ln \ w^2 \\  \\  
\displaystyle=-\frac 1{2\pi }\int_\Omega  A_\nu \ A_\nu\  d^2x\ \
 -\ \frac 1{4\pi }\ \ln \ w^2\int_{\partial \Omega }A_{\nu }\ dx_\nu . 
\end{array} \end{equation} The first term is the integral, restricted to
the region $\Omega $, of
 the same density appearing in the well known case of the whole plane
\cite {Schwinger}. The second term is well-defined for every $w\neq 0$,
and vanishes for a null total flux, $\Phi =0$. For $w=0$, $b_0$ in
(\ref{b0}) does not define an elliptic boundary problem.  It is also
interesting to notice that this term vanishes in the case of
 MIT bag boundary conditions, i.e., $w=\pm 1.$ 

This calculation is to be compared with the case of the compactified plane
\cite{Annals}, where the determinant can be expressed in terms of just the
kernel of the $z$-power of the operator analytically extended to $z=0$,
which is a local quantity. The presence of boundaries makes the evaluation
more involved, since even in simple cases as the present (or the half
plane treated in \cite{Jour.2}), the knowledge of the Green function of
the problem is needed.

\bigskip \bigskip { \bf \it Acknowledgments}.  We are grateful to R.T.
Seeley for useful comments.  \bigskip \bigskip

\bigskip
\eject

\end{document}